\documentclass[journal=jctcce,manuscript=article]{achemso}

\usepackage{rotating}
\usepackage{graphicx}
\usepackage{caption}
\usepackage{xcolor}
\usepackage{multirow}
\usepackage{multicol}
\usepackage{algorithm}
\usepackage{algorithmic}
\usepackage{amsmath}
\usepackage{amssymb}
\usepackage{bm}
\usepackage{ulem}
\usepackage{appendix}

\newcommand{\kan}[1]{{\color{violet}{#1 }}}

\newcommand{\added}[1]{\textcolor{black}{#1}}
\newcommand\COMMENTED[1] {}
\usepackage[version=3]{mhchem} % Formula subscripts using \ce{}

\author{Zhi-Yu Xiao}
\affiliation{%
Institute of Physics, Chinese Academy of Sciences, P.O. Box 603, Beijing 100190, China
}%
\email{zxiaoMain@outlook.com}
\altaffiliation{These authors contributed equally to this manuscript.}
\author{Bowen Kan}
\affiliation[ICT]
{Institute of Computing Technology, Chinese Academy of Sciences, Beijing 100190, China}
\alsoaffiliation[UCAS]
{University of Chinese Academy of Sciences, Beijing 101408, China}
\altaffiliation{These authors contributed equally to this manuscript.}
\author{Huan Ma}
\affiliation[USTC]
{State Key Laboratory of Precision and Intelligent Chemistry, University of Science and Technology of China, Hefei 230026, China}
\altaffiliation{These authors contributed equally to this manuscript.}

\author{Bowen Zhao}
\affiliation[USTC]
{State Key Laboratory of Precision and Intelligent Chemistry, University of Science and Technology of China, Hefei 230026, China}
\altaffiliation{These authors contributed equally to this manuscript.}

\author{Honghui Shang}
\email{shanghui.ustc@gmail.com}
\affiliation[USTC]
{State Key Laboratory of Precision and Intelligent Chemistry, University of Science and Technology of China, Hefei 230026, China}

\title[NNQS-AFQMC]
{NNQS-AFQMC: Neural network quantum states enhanced fermionic quantum Monte Carlo}

\abbreviations{NNQS, Transformer}

\begin{document}

\begin{tocentry}
	
	%\begin{figure}[htbp]
	\centering
	\includegraphics[width=6.6cm]{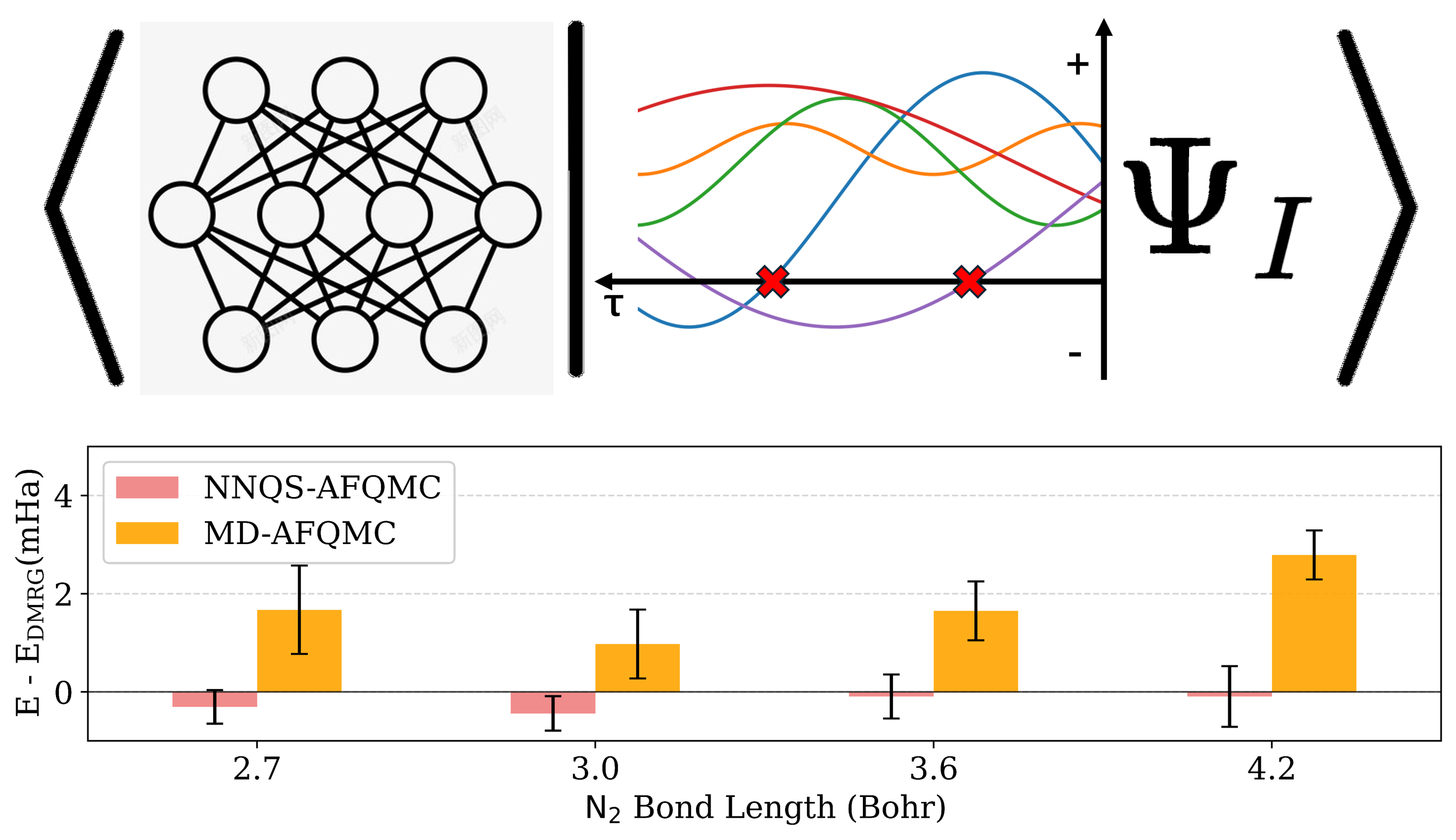}
	%    \label{fig:toc}
	%\end{figure}
	
	%Some journals require a graphical entry for the Table of Contents.
	%This should be laid out ``print ready'' so that the sizing of the
	
	%Inside the \texttt{tocentry} environment, the font used is Helvetica
	%8\,pt, as required by \emph{Journal of the American Chemical
	
	%The surrounding frame is 9\,cm by 3.5\,cm, which is the maximum
	%permitted for  \emph{Journal of the American Chemical Society}
	%graphical table of content entries. The box will not resize if the
	%content is too big: instead it will overflow the edge of the box.
	
	%This box and the associated title will always be printed on a
	%separate page at the end of the document.
	
\end{tocentry}

\begin{abstract}
We introduce an efficient approach to implement neural network quantum states (NNQS) as trial wavefunctions in auxiliary-field quantum Monte Carlo (AFQMC). NNQS are a recently developed class of variational ansätze capable of flexibly representing many-body wavefunctions, though they often incur a high computational cost during optimization. AFQMC, on the other hand, is a powerful stochastic projector approach for ground-state calculations, but it normally requires an approximate constraint via a trial wavefunction or trial density matrix, whose quality affects the accuracy. \added{Recently it has been shown (Xiao et al, arXiv2505.18519)  that a broad class of highly correlated wave-functions can be integrated into AFQMC through stochastic sampling techniques. In this work, we apply this approach and present a direct integration of NNQS with AFQMC,} allowing NNQS to serve as high-quality trial wavefunctions for AFQMC with manageable computational cost. We test the NNQS-AFQMC method on the challenging nitrogen molecule (N$_2$) at stretched geometries. Our results demonstrate that AFQMC with an NNQS trial wavefunction can attain near-exact total energies, highlighting the potential of AFQMC with NNQS to overcome longstanding challenges in strongly correlated electronic structure calculations. We also outline future research directions for improving this promising methodology.
\end{abstract}

\section{Introduction}
The exploration of interacting quantum many-body systems stands as a pivotal challenge that transcends multiple scientific frontiers, encompassing condensed matter physics, nuclear physics, cold atoms physics, quantum chemistry, and materials science. The complexity is deeply rooted in its intricate and non-trivial interplay among various degrees of freedom, which gives rise to various phenomena that are difficult to understand and predict. Over the last decades, remarkable progress has been made in theoretical frameworks and computational algorithms. However, each of these methods has its own limitations and is typically optimized for particular types of systems or physical regimes. No universal approach exists at present that can fully address the complexity of interacting quantum systems while providing systematically accurate results across a wide range of many-body models and materials. The continuous development of more general, computationally efficient, and highly accurate numerical techniques is essential. Such advancements would enable a deeper understanding of fundamental quantum phenomena and facilitate the design and discovery of novel quantum materials with potential applications in areas such as high-temperature superconductivity, quantum computing, and quantum information science.

In recent years, significant advancements in numerical algorithms within quantum chemistry have vividly demonstrated a pronounced trend towards interdisciplinary collaboration, particularly with the integration of artificial intelligence, a paradigm often referred to as ``AI for science''.  The application of neural networks in quantum chemistry has emerged as a rapidly evolving frontier, especially since Carleo and Troyer introduced the neural network quantum state (NNQS) method in 2017~\cite{CarleoTroyer2017}, which leverages neural networks to represent quantum wavefunctions and has shown significant promise in accurately describing complex quantum many-body systems. Their approach represents wavefunctions using restricted Boltzmann machines (RBMs) optimized through variational Monte Carlo (VMC), demonstrating remarkable effectiveness in capturing quantum system complexity. The NNQS methodology has since undergone rapid development and extension~\cite{ChooCarleo2019,SharirShashua2020,SchmittHeyl2020,YuanDeng2021,ZhaoLiang2022,MorenoStokes2022,ZhaoVeerapaneni2022}. Initial applications to molecular systems showed promising results, with RBMs achieving higher accuracy than CCSD(T) in various cases~\cite{ChooCarleo2020}. Since then, a variety of neural-network-based wavefunction ansätze have been developed for both model systems and real molecules, including autoregressive networks and others, achieving impressive accuracy in many cases~\cite{BarrettLvovsky2022, Shang2023, Wu2023_SC,lixiang_2024,HermannNoe2020,PfauFoulkes2020,sobral2024physicsinformedtransformerselectronicquantum}. Though continuous achievements have been made, the NNQS approach still suffers from the complexity of many-body problems. In particular, for large-scale strongly correlated systems, the computational cost remains prohibitively high. The training of neural networks requires extensive sampling and optimization; as the system size increases, the wavefunction complexity grows significantly, causing difficulties such as slower convergence, increased risk of overfitting, and prolonged training times. It is urgent to explore new algorithmic frameworks that go beyond the traditional variational ansatz and investigate hybrid approaches that combine neural networks with other quantum chemistry methods. 

Meanwhile, in quantum chemistry, auxiliary-field quantum Monte Carlo (AFQMC) has emerged as a powerful fermionic quantum Monte Carlo method in the study of many-electron systems. AFQMC is a projector-based algorithm, employs Monte Carlo sampling to simulate imaginary-time projection, and the infamous sign problem is mitigated by introducing trial wavefunctions to constrain its sampling process. AFQMC with constraint has enabled a wide range of applications where sign-problem-free computations are not feasible. These include strongly correlated models in condensed matter physics \cite{Bo_Xiao_Stripe_Hubbard, Hubbard_SC_HAOXU}, bulk solids \cite{Mario_hydrogen_chain,BM_model}, and quantum chemistry \cite{WIREs_Mario,Joonho_chemistry}. In practice, the accuracy of AFQMC depends on the quality of the trial wavefunction in characterizing the ground state.  In standard AFQMC implementations, Slater determinant wavefunctions and their variants are typically used as trial wavefunctions. These variants encompass orthogonal or non-orthogonal linear combinations of Slater determinants \cite{Hao_Some_recent_developments, WIREs_Mario, Morales_MD_AFQMC}, symmetry-projected mean-field wavefunctions \cite{Hao_symmetry}, projected Bardeen–Cooper–Schrieffer (BCS) states \cite{PRA_2011_2DFGs_Carlson, Vitali_Calculating_2019}, and pseudo-BCS states \cite{Zxiao_2020}. In benchmark studies of correlated fermion systems, this framework has shown good systematicity \cite{Hubbard_benchmark_2015, Joonho_chemistry, Mario_hydrogen_chain, simons_material_2020}. For strongly correlated systems, more sophisticated trial wavefunctions are required, and their implementation to AFQMC relies on configuration interaction (CI) approaches \cite{transition_metals_Shee, Hao_Some_recent_developments, SHCI_AFQMC, Ankit_CISD,sukurma2025self}. In extended systems, increasing the complexity of the trial wavefunction using a configuration interaction approach fails to achieve size-consistency, which restricts the broad application of AFQMC. The search for appropriate trial wavefunctions and their efficient implementation has long been a key challenge for AFQMC algorithms.

In this work, we combine the strengths of NNQS and AFQMC by implementing NNQS trial wave functions in AFQMC. Recent breakthroughs in quantum Monte Carlo methodology \cite{xiao2025implementingadvancedtrialwave} have established a novel framework for incorporating many-body trial wavefunctions with AFQMC. Building upon this foundational work, we demonstrate an approach to integrating sophisticated NNQS trial wavefunctions with AFQMC. Our method leverages the natural CI representation in most NNQS parameterizations, thus enabling a direct and efficient coupling between these two powerful computational paradigms. We demonstrate that this NNQS-enhanced AFQMC method yields highly accurate results for the prototypical strongly correlated N$_2$ molecule, achieving nearly exactness. The remainder of this paper is organized as follows. In Sec~.\ref{sec:Prilimilaries}, we provide a concise overview of NNQS and AFQMC methodologies. In Sec~.\ref{sec:methods_NNQS_AFQMC} we elaborate on the detailed integration of NNQS with AFQMC. In Sec~.\ref{sec:Results}, we present numerical results for N$_2$, highlighting the accuracy and efficiency of the combined approach. In Sec~.\ref{sec:Summary}, we conclude our work and discuss possible developments.

\section{\label{sec:Prilimilaries} Preliminaries}

For general quantum chemistry problems, the \textit{ab initio} Hamiltonian can be written as below
\begin{equation}
\hat{H}=\sum^N_{ij} \sum_{\sigma} h_{ij}a^\dagger_{i\sigma} a_{j\sigma} + \frac{1}{2}\sum_{ijkl}^N\sum_{\sigma \rho}V_{ijkl}a^\dagger_{i\sigma}a^\dagger_{j\rho}a_{k\rho}a_{l\sigma}
\label{eq:H}
\end{equation}
where $N$ denotes the number of basis involved, $ijkl$ and $\sigma \rho$ specify the index of basis and spin, respectively. $h_{ij}$ and $V_{ijkl}$ can be explicitly determined based on the chosen basis. 
% \mh{In our implementation, all Hamiltonians are defined on canonical Hartree-Fock orbitals.}

In the following, we introduce NNQS and AFQMC that are applied to evaluate the ground state of such an \textit{ab initio} Hamiltonian in Sec.~\ref{sec:methods_NNQS} and Sec.~\ref{sec:methods_AFQMC} respectively. Though these two methods rely on different principles, we will show in Sec.~\ref{sec:methods_NNQS_AFQMC} that they can be integrated to provide highly accurate results in a scaled way.

\subsection{Neural Network Quantum State}
\label{sec:methods_NNQS}
In quantum chemistry, the many-electron wavefunction can be expressed in second-quantized form as a linear combination of Slater determinant basis states (configurations):

\begin{equation}
|\Psi\rangle =\sum_{\mathbf{x}} \Psi(\mathbf{x})|\mathbf{x} \rangle
% \label{eq:NNQS}
\end{equation}
where $|x\rangle = |x_1, x_2, \dots, x_N\rangle$ represents a specific occupation configuration of $N$ electrons (with $x_i \in \{0,1\}$ indicating whether spin-orbital $i$ is occupied). The coefficient $\Psi(\mathbf{x})$ is generally complex and can be written as:
 \begin{equation}
\Psi(\mathbf{x}) = |\Psi(\mathbf{x})| e^{i\phi(\mathbf{x})}
\end{equation}
where $|\Psi(x)|$ and $\phi(x)$ are the amplitude and phase of the configuration $\mathbf{x}$. In our approach, we employ \textbf{QiankunNet}, a heterogeneous neural network, to represent the quantum state coefficients~\cite{Shang2023, Wu2023_SC}.  For the amplitude, we use a Transformer decoder architecture that processes an input configuration $\mathbf{x}$ through multiple self-attention layers. Formally, we embed the electron configuration (occupations and orbital positions) into an initial hidden representation $h^0$ using learnable embedding matrices $W_e$ and $W_p$. This representation is then passed through $n$ stacked Transformer decoder layers:
\begin{equation}
\begin{aligned}
h_0 &= XW_e+W_p \\
h_j &= \text{Decoder}(h_{j-1}), \quad j\in [1,n] \\
|\Psi_{\vec{\theta}}(\mathbf{x})|^2 &= \text{softmax}(h_nW_{\text{head}})
\label{eq:NNQS}
\end{aligned}
\end{equation}
where $W_{\text{head}}$ is a final linear projection, the softmax ensures the outputs form a normalized probability distribution and the network’s parameters $\vec{\theta}=(h_n, W_{\text{head}})$ are to be optimized. This yields $|\Psi_{\vec{\theta}}(x)|$ (the amplitude) for each configuration $\mathbf{x}$. The phase $\phi_{\vec{\theta}}(x)$ is generated by a separate multilayer perceptron (MLP) that takes either the same embedded input or the output of one of the decoder layers and produces $\phi_{\vec{\theta}}(x)$. By splitting amplitude and phase into two networks, we allow greater flexibility in capturing the sign structure of the wavefunction.

Training an NNQS to represent a ground-state wavefunction typically involves variational Monte Carlo (VMC), where the network’s parameters $\vec{\theta}$ are optimized to minimize the energy expectation value:
In this formulation, the system's energy expectation value becomes a parametric function $E(\vec{\theta})$ dependent on the variational parameters $\vec{\theta}$ :

\begin{equation}
\begin{split}
    E(\vec{\theta})
= \frac{\langle \Psi | H | \Psi \rangle}
       {\langle \Psi | \Psi \rangle}
= \frac{\displaystyle \sum_{\mathbf{x}} E_{loc}(\mathbf{x})
       \,\bigl|\Psi_{\vec{\theta}}(\mathbf{x})\bigr|^2}
     {\displaystyle \sum_{\mathbf{x}} \bigl|\Psi_{\vec{\theta}}(\mathbf{x})\bigr|^2}
    \label{Energy}
\end{split}
\end{equation}
where
\begin{equation}
E_{loc}(\mathbf{x})
= \sum_{\mathbf{x}'} H_{\mathbf{x} \mathbf{x}'} \,\Psi_{\vec{\theta}}(\mathbf{x}')/{\Psi_{\vec{\theta}}(\mathbf{x})}.
    \label{local_energy}
\end{equation}
and $H_{\mathbf{x} \mathbf{x}'} = \langle \mathbf{x} | H| \mathbf{x}' \rangle$ is the Hamiltonian matrix element. Direct computation of Eq.\ref{Energy} is typically intractable because the number of possible configurations grows exponentially with system size. To overcome this limitation, we evaluate the energy as an expectation over the probability distribution $p_{\vec{\theta}}(\mathbf{x}) = \bigl|\Psi_{\vec{\theta}}(\mathbf{x})\bigr|^2$, which is done by Monte Carlo sampling over configurations $\mathbf{x}$. The estimation of energy average local energy measurements over these sampled configurations $\left\{ \mathbf{x}^i \right\}_{i=1}^{N_s}$:
\begin{equation}
    \widetilde{E}\bigl(\vec{\theta}\bigr)
= \frac{1}{N_s} \sum_{i=1}^{N_s} E_{loc}\bigl(\mathbf{x}^i\bigr).
    \label{energy expectation}
\end{equation}

The gradient-driven optimization approaches often yield significantly better computational performance than their derivative-free counterparts. The gradients of Eq.~\ref{Energy} can be efficiently computed via automatic differentiation by leveraging the sampled configurations:
\begin{equation}
\nabla_{\vec{\theta}} \,\tilde{E}
= 2\,\mathrm{Re}\,
\Bigl(\mathbb{E}_{p}\!\Bigl[
  \bigl(E_{loc}(\mathbf{x}) - \mathbb{E}_{p}[\,E_{loc}(\mathbf{x})\,]\bigr)\,
  \nabla_{\vec{\theta}} \ln\!\bigl(\Psi_{\vec{\theta}}^*(\mathbf{x})\bigr)
\Bigr]\Bigr)
    \label{gradient}
\end{equation}
where the gradient estimator $\nabla_{\vec{\theta}} \,\tilde{E}$, serving as an approximation to the exact energy gradient $\nabla_{\vec{\theta}} \,{E}$, cooperates with optimizer to refine the parameter $\vec{\theta}$, thereby constituting a complete iteration in the VMC algorithm.

\subsection{Auxiliary-Field Quantum Monte Carlo (AFQMC)}
\label{sec:methods_AFQMC}

The application of AFQMC to real materials start from the ``Monte Carlo form" \cite{Hao_Some_recent_developments}:
\begin{equation}
\hat{H} = \sum^N_{ij} \sum_{\sigma} h_{ij}a^\dagger_{i\sigma} a_{j\sigma} +\frac{1}{2}\sum^\Gamma_\gamma \hat{L}_\gamma ^2\,,
\label{eq:H_MC}
\end{equation}
where $\hat{L}_\gamma$ denote sets of one-body operators, whose matrix elements 
depend on details of the ``factorization" transformation
(a modified Cholesky \cite{Cholesky_dec}, density-fitting \cite{GPU_AFQMC_James}, etc) but 
are explicitly specified from $\{V_{ijkl}\}$ that reproduces Eq.~\ref{eq:H}, and $\Gamma\sim {\mathcal O}(10)\,N$.  

The ground-state calculation is performed through the imaginary-time projection $e^{- \beta\hat{H} }$ with sufficiently long imaginary time $\beta$ on an initial state $|\Psi_{I}\rangle$ that has a non-zero overlap with the ground state. This is achieved by employing the Suzuki-Trotter decomposition \cite{Suzuki,Trotter}, which divides the imaginary-time projector into small time slices:
\begin{equation}
|\Psi_{GS}\rangle \propto \lim_{\beta \to \infty} e^{- \beta\hat{H} }  = (e^{-\tau \hat{H} })^n|\Psi_{I}\rangle\,, 
\label{eq:imaginary_time_evolve}
\end{equation}
where $n= \beta/\tau $ denotes the number of time slices. The length of a time slice $\tau$ must be sufficiently small to minimize the errors from commutators (compared to MC statistical errors). The Hubbard-Stratonovich transformation \cite{HS_transformation_H, HS_transformation_S} is applied to decouple the short-time projection in each slice:
\begin{equation}
e^{-\tau \hat{H} }\approx  \int\mathrm{d}\textbf{y}\, p(\textbf{y})\,\hat{B}(\textbf{y})\,,
\label{eq:HS}
\end{equation}
where $p(\textbf{y})$ is a probability density function, which can be either discrete or continuous. For example, a Gaussian distribution is commonly used:
\begin{equation}
p(\textbf{y})= \prod_\gamma \frac{1}{\sqrt{2\pi}}e^{-y_\gamma^2/2}\propto e^{-\textbf{y}\cdot\textbf{y}/2},
\end{equation}
and the one-body propagator $\hat{B}(\textbf{y})$ is given by \cite{Cholesky_dec}
\begin{equation}
\hat{B}(\textbf{y})=e^{-\tau\hat{T}/2}e^{\sum_\gamma y_\gamma\sqrt{-\tau}\hat{L}_\gamma }e^{-\tau\hat{T}/2}.
\label{eq:HS-B}
\end{equation}
where $\textbf{y}$ denotes a series of $\{y_\gamma\}$ with $y_\gamma \in \mathbb{R}$ and $\textbf{y}\cdot\textbf{y}=\sum_{\gamma}y_\gamma^2$. Since the projection of these one-body operators $\hat{B}(\textbf{y})$ to a single determinant $|\phi\rangle$ is another single determinant: $|\phi'\rangle \propto \hat{B}(\textbf{y})|\phi\rangle$ and, without loss of generality, $|\Psi_{I}\rangle$ can be chosen as a sum of single determinants, the sampling of auxiliary fields $\textbf{y}$ is mapped into random walks in the manifold of Slater determinants along the direction of imaginary time evolution \cite{lecturenotes-2019}. 

Fermions are described by anti-symmetric wavefunctions, which leads to the infamous phase problem in Monte Carlo sampling methods. In AFQMC, the phase problem arises during imaginary time evolution when walkers randomly become perpendicular to the ground state, which contributes only statistical noise to the ground-state estimation. The number of such perpendicular determinants typically grows exponentially with projection time, causing the Monte Carlo signal to be overwhelmed by statistical fluctuations. \added{To be more clear, once walkers become perpendicular to the ground state: $\langle \Psi_{GS}|\phi\rangle=0$, the following imaginary time projection on walker $|\phi\rangle$ contributes nothing to the whole estimation of the ground state as $\langle \Psi_{GS}|e^{-\hat{H}\beta}|\phi\rangle=0$, leading to the exponential growth of statistical fluctuations.
Therefore,} eliminating these determinants that are perpendicular to the ground state is an exact condition which removes the sign or phase problem \cite{QMC_Zhang_Constrained_1997, AFQMC_Zhang-Krakauer-2003-PRL}. 

To resolve the phase problem, ground state AFQMC utilizes a trial wavefunction $\langle \Psi_T|$ to constrain the random walk paths. This is accomplished by introducing a dynamic shift $\overline{\textbf{y}}$ in the integration over auxiliary fields in Equation (\ref{eq:HS}):
\begin{equation}
e^{-\tau \hat{H} }= \int\mathrm{d}\textbf{y}\, p(\textbf{y}- \overline{\textbf{y}})\,\hat{B}(\textbf{y}- \overline{\textbf{y}})\,.
\label{eq:HS-shifted}
\end{equation}
With this equation, each propagation step can be executed:
\begin{equation}
e^{-\tau \hat{H} }\,
\sum_k W^{i}_k \frac{|\phi^{i}_k \rangle}{\langle \Psi_T|\phi^{i}_k \rangle}
\rightarrow
\sum_k W^{i+1}_k \frac{|\phi^{i+1}_k \rangle}{\langle \Psi_T|\phi^{i+1}_k \rangle}\,.
\label{eq:prop-imp}
\end{equation}
This advancement moves the walker according to:
\begin{equation}
\begin{aligned}
\hat{B}(\textbf{y} - \overline{\textbf{y}}_k^i)| \phi_k^i \rangle
\rightarrow | \phi_k^{i+1} \rangle\,,
\label{eq:walker_prop}
\end{aligned}
\end{equation}
and assigns a new weight:
\begin{equation}
\begin{aligned}
I(\textbf{y}, \overline{\textbf{y}}_k^i , \phi_k^i)\,W_k^i \rightarrow W_k^{i+1}
\label{eq:weight_prop}
\end{aligned}
\end{equation}
where the importance function $I(\textbf{y}, \overline{\textbf{y}}_k^i , \phi_k^i)$ is defined as:
\begin{equation}
\begin{aligned}
I(\textbf{y}, \overline{\textbf{y}}^{i}_k, \phi^{i}_k) = \frac{p(\textbf{y}-\overline{\textbf{y}}^{i}_k)}{p(\textbf{y})}\frac{\langle \Psi_T| \hat{B}(\textbf{y} - \overline{\textbf{y}}_k^i)| \phi^{i}_k \rangle }{\langle \Psi_T|\phi^{i}_k \rangle }\,.
\end{aligned}
\label{eq:importance_function}
\end{equation}
The shift $\overline{\textbf{y}}_k^i$ is selected to minimize the fluctuations in the weights. For small $\tau$, the optimal choice for each component is given by \cite{AFQMC_shift}:
\begin{equation}
\begin{aligned}
\overline{\mathbf{y}}^{\mathbf{i}}_{k,\boldsymbol{\gamma}} = -\sqrt{\tau}\frac{\langle \Psi_T|\hat{L}_{\boldsymbol{\gamma}}|\phi^{\mathbf{i}}_{k} \rangle}{\langle \Psi_T| \phi^{\mathbf{i}}_{k}\rangle}\,.
\end{aligned}
\label{eq:shift1}
\end{equation}
When walkers are propagated through the above procedure to the $n_{\rm th}$ step and reach equilibrium (i.e., when $\beta=n_{\rm eq}\tau$ is sufficiently large to reach the ground state in Eq.~\ref{eq:imaginary_time_evolve} within the desired statistical accuracy), the ground - state energy can be measured using the mixed estimator \cite{AFQMC_Zhang-Krakauer-2003-PRL,WIREs_Mario} for all $n > n_{\rm eq}$:
\begin{equation}
\frac{\langle \Psi_T |\hat{H} |\Psi^n \rangle}{\langle \Psi_T |\Psi^n \rangle}=\frac{\sum_{k} W^{n}_k\frac{\langle \Psi_T |\hat{H} |\phi^{n}_k\rangle}{\langle \Psi_T|\phi^{n}_k\rangle}}{\sum_{k } W^{n}_k}.
\label{eq:mixed_estimator}
\end{equation}
The importance sampling formalism, combined with the force bias, automatically enforces the constrained path approximation \cite{QMC_Zhang_Constrained_1997,lecturenotes-2019} in the presence of a sign problem. For \textit{ab initio} Hamiltonians, an additional phaseless approximation \cite{AFQMC_Zhang-Krakauer-2003-PRL,lecturenotes-2019,Hao_Some_recent_developments} is applied. This approximation projects the weight of each Slater determinant onto the real axis in the complex plane \cite{Hao_Some_recent_developments}:
\begin{equation}
I{_\textup{ph}}(\textbf{y}, \overline{\textbf{y}}^{i}_k, \phi^{i}_k) = \bigl|I(\textbf{y}, \overline{\textbf{y}}^{i}_k, \phi^{i}_k)\bigr|,* \textup{max}(0, \textup{cos}(\Delta \theta))\
\label{eq:importance_function_CP}
\end{equation}
where
\begin{equation}
\Delta\theta=\textup{Arg}\frac{\langle \Psi_T| \phi^{i+1}_k \rangle }{\langle \Psi_T|\phi^{i}_k \rangle }.
\end{equation}
The specific implementation details can vary. For example, there are local energy versus hybrid methods \cite{hybrid_phaseless}, methods that remove the absolute value in Equation (\ref{eq:importance_function_CP}) by maintaining the overall phase during long projection times \cite{AFQMC_Zhang-Krakauer-2003-PRL, AFQMC_PathRestoration}, or techniques such as cosine projection, the half-plane method, or others used to eliminate the finite density at the origin in the complex plane \cite{Hao_Some_recent_developments}. When all auxiliary fields are real (such as in the case of Hubbard interactions), the phaseless formalism simplifies to the constrained path approach \cite{QMC_Zhang_Constrained_1997}, with $\Delta\theta = 0$. The zero values in the importance function prevent the random walk from encountering the sign problem

The implementation of constraints introduces a systematic bias that depends on how well the trial wavefunction $\langle \Psi_T|$ can characterize the ground state. Extensive benchmark studies have demonstrated the high accuracy of AFQMC, both in models \cite{Hubbard_benchmark_2015, Hubbard_SC_HAOXU} and real materials \cite{simons_material_2020, Mario_hydrogen_chain, Joonho_chemistry}. As we discussed above, once the overlap between the trial wavefunction and the general Slater determinant can be evaluated, the implementation of the trial wavefunction to AFQMC is straightforward, and the remaining difficulty is how efficiently this evaluation can be. A wide variety of trial wavefunctions have been developed for AFQMC, including single determinants \cite{QMC_Zhang_Constrained_1997, AFQMC_Zhang-Krakauer-2003-PRL}, multi-Slater determinants \cite{Joonho_chemistry}, Bardeen–Cooper–Schrieffer (BCS) states \cite{Ettore_BCS, Hao_HFB}, tensor-network states \cite{MPS_AFQMC} and CI excited states \cite{Morales_MD_AFQMC, Hao_Some_recent_developments, SHCI_AFQMC, Ankit_CISD}. However, these wavefunctions often suffer from limited expressiveness or the absence of robust optimization tools.  

\section{Integration of NNQS with AFQMC}
\label{sec:methods_NNQS_AFQMC}
Recent advancement \cite{xiao2025implementingadvancedtrialwave} has shown that a broad class of many-body trial wavefunctions can be effectively implemented in AFQMC via stochastic sampling. Leveraging such progress, we introduce a novel integration of NNQS as trial wavefunctions with AFQMC. In this section, we provide a concise summary of the work in Ref \cite{xiao2025implementingadvancedtrialwave} to introduce the stochastic sampling approach and elaborate on details in its application to NNQS trial wavefunctions. 

The main conclusion in Ref\cite{xiao2025implementingadvancedtrialwave} is that an efficient implementation of the many-body trial wavefunction $\langle \Psi_T|$ to AFQMC by stochastic sampling can be devised if there is an integral representation of such a trial wavefunction:
\begin{equation}
\langle\Psi_T| = \int d\mathbf{x} \langle \Psi_T|\mathbf{x}\rangle \langle\mathbf{x}| .
\label{eq:general_trial}
\end{equation}
where the variable $\mathbf{x}$ parameterizes a family of states $| \mathbf{x}\rangle$ and its overlap $\langle \mathbf{x}|\phi\rangle$ with Slater determinant $|\phi\rangle$ can be evaluated effectively. This can be understood by noticing that trial wavefunction $\langle\Psi_T|$ only appears in the evaluation of overlap ratios and “local” quantities—such as force biases and local energies—during the AFQMC procedure detailed in Sec.~\ref{sec:methods_AFQMC}. Once a specific integration representation is defined, stochastic sampling enables accurate estimation of these quantities. In Ref\cite{xiao2025implementingadvancedtrialwave}, authors demonstrated this approach by integrating Variational AFQMC (VAFQMC) trial wavefunctions with AFQMC, based on an integration representation formulated in the auxiliary field space. This idea can be generalized to NNQS trial wavefunctions, based on an integration representation formulated in the configuration space. 

For NNQS, there is a natural integral representation in the configuration space as discussed in Sec.~\ref{sec:methods_NNQS}. The Eq.~\ref{eq:general_trial} reduces to the integration representation of NNQS trial wavefunction :
\begin{equation}
\langle\Psi_T| = \int d\mathbf{x} \Psi(\mathbf{x}) \langle \mathbf{x}| .
\label{eq:NNQS_trial}
\end{equation}
with the variable $\mathbf{x}$ denotes the configuration within the configuration space, and the coefficient $\Psi(\mathbf{x})=\langle \Psi_T|\mathbf{x}\rangle$ is determined in the NNQS trial wavefunction. The evaluation of overlap $\langle \mathbf{x}|\phi\rangle$ between the configuration $\mathbf{x}$ and the Slater determinant $|\phi\rangle$ is straightforward\cite{Hao_Some_recent_developments}. The coefficient $\Psi(\mathbf{x})$ is systematically optimized in NNQS to approximate the ground state. As detailed in Sec.~\ref{sec:methods_NNQS}, once the NNQS parameters are optimized, these coefficients can be efficiently obtained, enabling their integration with AFQMC.

\begin{figure}[htbp]
\includegraphics[scale = 0.6]{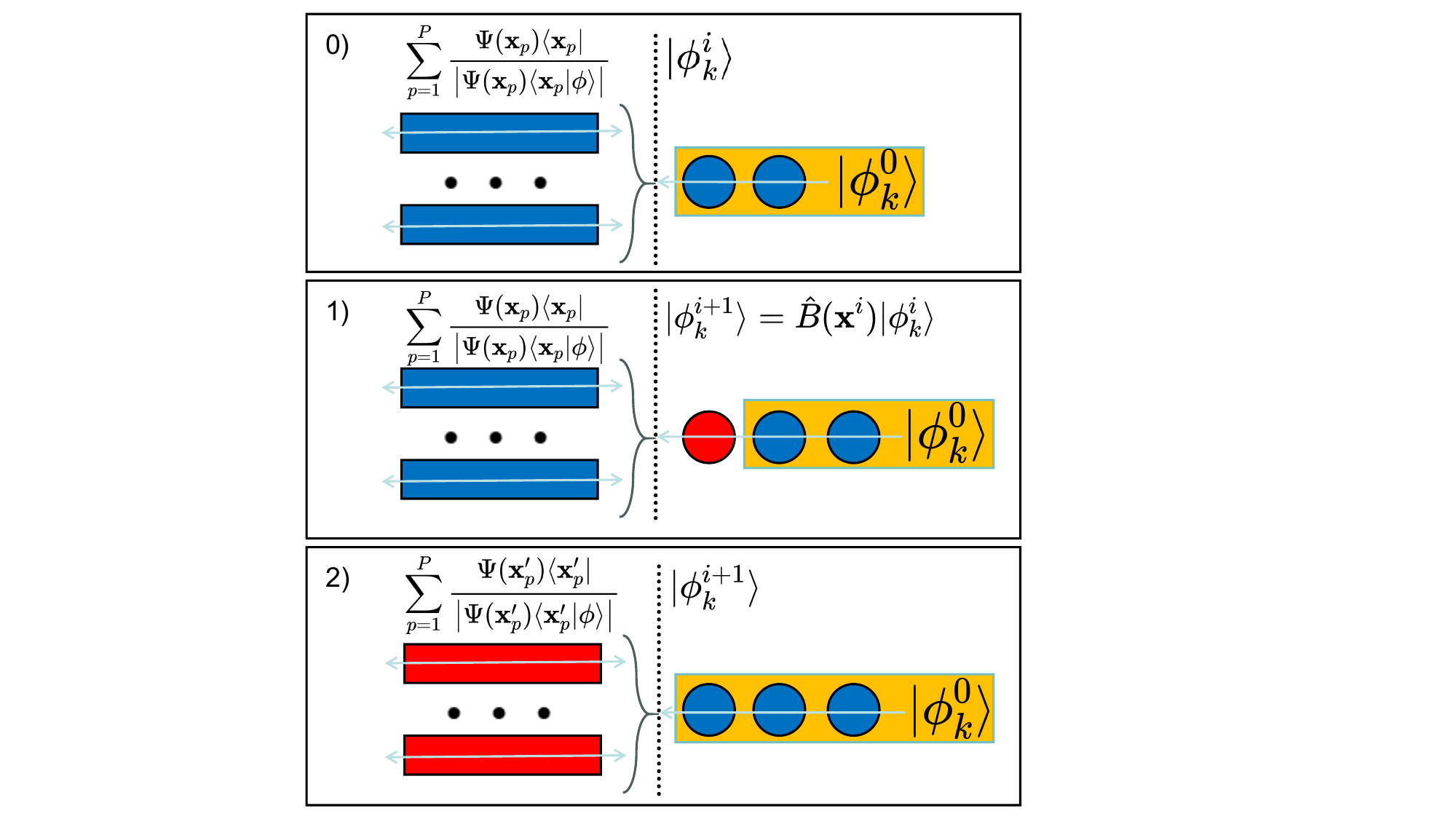} 
\caption{
Illustration of the integration of NNQS with AFQMC. One random AFQMC walker and its tethered MCMC sampled configurations are shown, during one leapfrog update step. Circle and rectangular denote one-body operators $\hat B(\textbf{y})$ and configurations $\langle \mathbf{x}_p|$, respectively. Blue/red indicates current/updated state in the MC. 
0). current state of the walker $|\phi^{i}_k \rangle$ and MCMC sampled configurations $\langle \bar \Psi_T|$ attached to it. 
1). the walker is propagated one step, $|\phi^i_k \rangle \rightarrow |\phi^{i+1}_k \rangle$ and its weight updated $W^i_k  \rightarrow W^{i+1}_k$,
using the current samples of  $\langle \bar \Psi_T|$ as importance function and constraint. 
2). MCMC sweeps are performed to update each configurations (each $p$, represented by one row of rectangular), 
$\mathbf{x}_{k,p}^i \rightarrow \mathbf{x}_{k,p}^{i+1}$, according to the walker's new position, $|\phi^{i+1}_k \rangle$. 
}
\label{Fig.AFQMC_Metro} 
\end{figure}

To allow a stochastic sampling of the trial wavefunction (i.e., Eq.~\ref{eq:NNQS_trial}) in AFQMC, the main modification to the standard AFQMC process is to reformulate the propagation of walkers in Eq.~\ref{eq:prop-imp}. Conceptually, this involves augmenting the standard AFQMC ensemble of random walkers $\{|\phi^{i}_k\rangle, W_k^i\}$ with an additional set of $P$ configurations $\{\mathbf{x}_{k,p}^i\}$ ($p=1,2,\cdots, P$) associated with each walker. Here, $\{\mathbf{x}_{k,p}^i\}$ are sampled from the integration space defined in Eq.~\ref{eq:NNQS_trial}, allowing the trial wavefunction to be stochastically approximated. For notational clarity, we suppress the indices $k$ and $i$ in the following discussion, denoting the current walker state as $|\phi\rangle$ with weight $W$ and associated sampled configurations $\mathbf{x}_p$, and the propagated state as $|\phi'\rangle$ with weight $W'$ and $\mathbf{x}'_p$ (analogous to the blue and red objects in Fig.~\ref{Fig.AFQMC_Metro}). The many-body trial wavefunction can then be estimated via:

\begin{equation}
\langle \Psi_{T}| \doteq \langle \bar \Psi_{T}|
=\frac{{\mathcal N}(\phi)}{P} \sum_{p=1}^P
\frac{\Psi(\mathbf{x}_p)\langle \mathbf{x}_p |}
{\big| \Psi(\mathbf{x}_p)\langle \mathbf{x}_p|\phi\rangle \big|},
\label{eq:Metropolis_sampled_trial}
\end{equation}
where the normalization factor
${\mathcal N}(\phi)\equiv
{\int \big| \Psi(\mathbf{x}_p)\langle \mathbf{x}_p|\phi\rangle \big|\,d\mathbf{x}}$
emerges naturally from the sampling procedure but does not need to be computed explicitly. The samples $\mathbf{x}_p$ are drawn from the probability distribution ${\mathcal P}(\mathbf{x}_p;\phi) = \big| \Psi(\mathbf{x}_p)\langle \mathbf{x}_p|\phi\rangle \big|/{\mathcal N}(\phi)$
ensuring that the ensemble average in Eq.~\ref{eq:Metropolis_sampled_trial} converges to the desired trial wavefunction, which is achieved by MCMC in this work.

Based on Eq.~\ref{eq:Metropolis_sampled_trial}, the overlap ratio can be estimated using the following formula:
\begin{equation}
\frac{\langle \bar \Psi_{T}|\phi' \rangle}{\langle \bar \Psi_{T}|\phi \rangle}
=
\frac{\sum_{p=1}^P
\frac{\langle \mathbf{x}_p |\phi'\rangle}{\langle \mathbf{x}_p |\phi\rangle}S(\mathbf{x}_p)}
{\sum_{p=1}^P S(\mathbf{x}_p)},
\label{eq:ratio-update-phi}
\end{equation}
where
\begin{equation}
S(\mathbf{x}_p)\equiv
\frac{\Psi(\mathbf{x})\langle \mathbf{x}_p |\phi\rangle}{\Big|\Psi(\mathbf{x})\langle \mathbf{x}_p |\phi\rangle\Big|}
\label{eq:sign-path}
\end{equation}
serves as a phase factor for each of the P configurations associated with the walker $|\phi\rangle$. Similarly, the local expectation of any observable $\hat O$ for the state $|\phi\rangle$ can be estimated as:
\begin{equation}
\frac{\langle \bar \Psi_{T}|\hat O |\phi \rangle}{\langle \bar \Psi_{T}|\phi \rangle}\doteq
\frac{\sum_{p=1}^P
\frac{\langle \mathbf{x}_p |\hat O|\phi\rangle}{\langle \mathbf{x}_p |\phi\rangle}S(\mathbf{x}_p)}
{\sum_{p=1}^P S(\mathbf{x}_p)}.
\label{eq:local-obs}
\end{equation}
The computed values from the above equations enable the calculation of the force bias described in Eq.~\ref{eq:shift1}, which is essential for completing the propagation of walkers as Eq.~\ref{eq:walker_prop}. The weight update from $W$ to $W'$ is achieved through the function $I_{\rm ph}$ defined in Eq.~\ref{eq:importance_function_CP}. With these fundamental components in place, the integration of NNQS into the AFQMC framework becomes feasible. 

The whole algorithm is performed in a leapfrog manner, that summarized in Fig.~\ref{Fig.AFQMC_Metro}. As presented, the standard AFQMC propagation is first performed to advance walker $\{|\phi\rangle, W\} \rightarrow \{|\phi'\rangle, W'\}$ according to $\langle \bar \Psi_{T}|$ with configurations $\{\mathbf{x}_p\}$ that sampled from the probability distribution ${\mathcal P}(\mathbf{x}'_p;\phi')$. Then, as walker is advanced to $|\phi'\rangle$, MCMC is adopted to update $\langle \bar \Psi_{T}| \rightarrow \langle \bar \Psi'_{T}|$ with newly sampled configurations $\{\mathbf{x}'_p\}$ according to the updated probability distribution ${\mathcal P}(\mathbf{x}'_p;\phi')$. One noticed ``hand-off'' occurs between two stochastic samples of the trial wavefunction. That is, considering the statistical fluctuation of Monte Carlo sampling, the update of configurations $\{\mathbf{x}'_p\}$ can leads to an extra ratio of overlaps being inserted before the beginning of the next AFQMC step: ($|\phi'\rangle \rightarrow |\phi''\rangle$):
\begin{equation}
R(\bar \Psi_T\to \bar\Psi_T';\phi')=\frac{\langle \bar \Psi_T'| \phi' \rangle }{\langle \bar \Psi_T|\phi'\rangle }\,,
\label{eq:AFQMC_Metro_correction}
\end{equation}
which is identity when the NNQS trial wavefunction is evaluated exactly by $\langle \bar \Psi_T|$. This statistical fluctuation can lead to a violation of the constraint condition, and we modify the weight of the walker $|\phi'\rangle$ by the cosine projection to correct for it:
\begin{equation}
W'\rightarrow W'\,\cdot \max(0,\cos(\Delta\theta_T))\,,
\label{eq:AFQMC_Metro_constraint_cos}
\end{equation}
where $\Delta\theta_T=\textup{Arg}[R(\bar \Psi_T\to \bar\Psi_T';\phi')]$. For a comprehensive understanding of the detailed algorithm in implementing NNQS, readers are referred to the APPENDIX.~\ref{sec:APPENDIX_summary}. 

It is important to emphasize that the phase factor $S(\mathbf{x}_p)$ appearing in Eqs.~\ref{eq:ratio-update-phi} and \ref{eq:local-obs} is not associated with the presence of the ``phase problem". \added{In other words, the value of $\frac{1}{P}\sum_{p=1}^P S(\mathbf{x}_p)$ won't reduce to zero during the propagation of AFQMC, if it is finite in the initialization of AFQMC.} Conceptually, this distinction arises because, in the limit of infinite samples ($P \rightarrow \infty$), the denominator $\frac{1}{P}\sum_{p=1}^P S(\mathbf{x}_p)$ in Eq.~\ref{eq:ratio-update-phi} converges to a finite positive value or such configurations are naturally removed by the constraint process, ensuring the stability of the algorithm. However, in practical implementations with finite $P$, statistical fluctuations in the sampled phase factors can introduce bias. Specifically, the approximation in Eq.~\ref{eq:AFQMC_Metro_constraint_cos} relies on the assumption that $R(\bar \Psi_T\to \bar\Psi_T';\phi')$ is close to identity. When $P$ is insufficient, this assumption breaks down, leading to systematic errors in the updates. These errors manifest as increased variance in the estimation and, in extreme cases, can induce numerical instability. \added{In this work, all calculations are performed under a significant $\frac{1}{P}\sum_{p=1}^P S(\mathbf{x}_p) \approx 1$.}

In the practical implementation, instead of directly invoking NNQS within the AFQMC framework (i.e., coefficients $\langle\Psi_T|\mathbf{x}_p\rangle$ are computed on-the-fly during AFQMC iterations), we obtain coefficients from a pre-computed dataset. Specifically, all configurations $\mathbf{x}_p$ and their corresponding coefficients $\langle\Psi_T|\mathbf{x}_p\rangle$ are precomputed and stored on disk. In this work, we saved all configurations in NNQS Monte Carlo estimation (Eq.~\ref{energy expectation}) as our precomputed dataset, and the number of stored configurations is around $O(10^5\sim 10^6)$. Such a choice of configurations is sufficient to characterize the essential features of the NNQS wavefunction. By doing so, the sampling of the NNQS trial wavefunction reduces to a sampling from the precomputed dataset. And the detailed MCMC process to sample from such datasets can be further optimized. Beyond the standard configuration space proposal strategy—where new configurations are generated via electron hopping/exchange from old configurations—we introduce a ``sorted labeling" strategy. Configurations are first sorted by the absolute value of their coefficients and assigned sequential labels. New states are then proposed by randomly selecting labels in the vicinity of the old label whose coefficients are comparable to each other. This construction yields a substantial improvement in MCMC acceptance ratio compared to standard configuration space proposals, particularly when the magnitude of coefficients is associated with the importance of configurations in the consist of NNQS wavefunction, which is a characteristic feature of NNQS wavefunctions. In this work, the ``sorted labeling" strategy is implemented in our MCMC proposing process to effectively integrate with NNQS trial wavefunctions. \added{More details about the ``sorted labeling" strategy is elaborated in APPENDIX.~\ref{sec:appendix_sorted_labeling}}

Our integration of NNQS with AFQMC scales linearly to the number of Metropolis samples ($P$), introducing a multiplicative prefactor to the cost of standard AFQMC with single-determinant trial wavefunctions. This prefactor is significantly smaller than $P$ by adopting an efficient Sherman-Morrison-Woodbury (SMW) algorithm \cite{Hao_Some_recent_developments, GPU_AFQMC_James} that accelerates the computation of the ratio $\frac{\langle \mathbf{x}'|\phi\rangle}{\langle \mathbf{x}|\phi\rangle}$ between ``adjacent" configurations $\mathbf{x}$ and $\mathbf{x}'$, \added{reducing the cost of MCMC procedures in updating configurations}. Conceptually, the computational cost of our approach with $P$ samples is comparable to previous AFQMC algorithms with $P$ configurations in Complete Active Space (CAS) trial wavefunctions \cite{Hao_Some_recent_developments, Joonho_chemistry}, which usually handles $P \sim O(10^3)$ configurations. One may notice that there is a naive implementation of NNQS with AFQMC by directly selecting configurations with the largest coefficients in NNQS as the trial wavefunction, which is similar to the implementation with CAS trial wavefunctions. However, this naive integration proves ineffective. The number of configurations needed to describe NNQS grows exponentially, and $O(10^3)$ configurations fall significantly short of fulfilling this requirement. To show this, a comparison between our implementation in this work and the naive implementation is presented in APPENDIX.\ref{sec:APPENDIX_MD-AFQMC}.

The computational bottleneck of our calculations resides in the training of NNQS, which is challenged by extensive sampling and prolonged optimization steps as the system size increases, especially for strongly correlated systems. To tackle this issue, instead of performing a single AFQMC calculation with a fully converged NNQS trial wavefunction, we conduct multiple AFQMC runs using a series of NNQS trial wavefunctions optimized with increased training steps. This allows us to monitor AFQMC energy convergence during the NNQS training process. Such integration leverages the observation that AFQMC energies converge well before the convergence of NNQS training, enabling high-accuracy results with significantly reduced computational overhead. As illustrated in Fig.~\ref{fig:diff_nnqs_wf}, this approach achieves near-exact results while reducing NNQS training steps overhead. These results highlight a major improvement in the efficiency of our integration.

\section{Results}
\label{sec:Results}

In this section, the performance and advantages of our method are examined through its application to $\mathrm{N}_2$. Given the challenge of accurately describing the triple-bond dissociation, $\mathrm{N}_2$ bond breaking serves as a typical system to benchmark numerical algorithms. We systematically investigated $\mathrm{N}_2$ at a series of increased bond lengths from 2.118 Bohr (equilibrium) to a stretched length of 4.2 Bohr, ranging from weakly and strongly correlated electronic regimes.
DMRG reference \cite{N2_bondbreaking_DMRG} and previous AFQMC results \cite{AFQMC_bondBreaking} are benchmarked with this work. In the following, we introduce the detailed workflow of our calculations with practical parameters, summaries of our main results, and then focus on two detailed but crucial observations in our integration of NNQS with AFQMC. Though our integration is general for any NNQS, in this work, all NNQS trial wave functions are generated from QiankunNet \cite{Shang2023,Wu2023_SC}.

Our calculations commences with Restricted Hartree-Fock (RHF) executed through PySCF\cite{pyscf_1, pyscf_2}. The outputs, including molecular orbitals and electron integrals, serve as inputs for QiankunNet and AFQMC, ensuring both methods are based on the same second-quantized Hamiltonian \added{defined on canonical Hartree-Fock orbitals} (Eq.~\ref{eq:H}). As proposed in Sec.~\ref{sec:methods_NNQS_AFQMC}, a series of AFQMC calculations are carried out with NNQS trial wavefunctions along the training process until AFQMC energies are converged within statistical error. \added{NNQS wavefunctions are optimized with an AdamW optimizer.} A consistent set of parameters is used for all AFQMC calculations: time slice $\tau = 0.01$, $2000$ steps for thermalization, and $200$ measurements taken at intervals of $25$ steps with $128$ walkers after the equilibrium. An increased number of configurations $P$ in AFQMC is chosen to sample the NNQS trial wavefunction, ensuring the convergence of our calculations. All calculations are executed on AMD EPYC 7763 Processors with 128 cores. QiankunNet is executed on a hybrid CPU-GPU architecture, utilizing an Intel Xeon Platinum 8480+ processor alongside NVIDIA A800 GPUs. The system operates on Ubuntu 22.04 LTS, with Python 3.11 and CUDA 12.2 for compatibility. PySCF is also deployed on the same hardware, maintaining consistent Python dependencies. 

\begin{figure}
    \centering
    \includegraphics[width=1.0\linewidth]{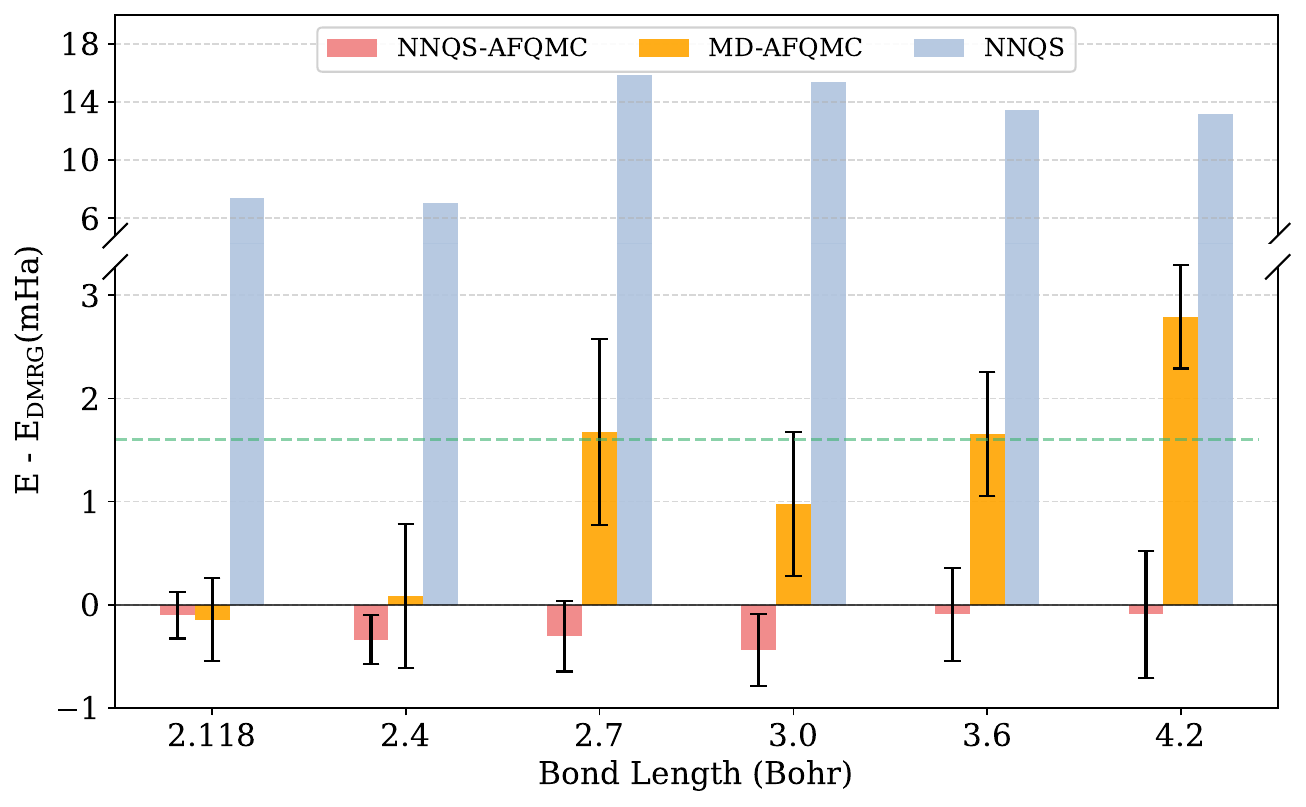}
    \caption{
    Deviations in the computed ground-state energy from NNQS and NNQS-AFQMC, for the N$_2$ molecule at different bond lengths, in the cc-pVDZ basis. 
    Energies are compared with near-exact DMRG results \cite{N2_bondbreaking_DMRG, AFQMC_bondBreaking}. Here, NNQS results do not stand for the optimized QiankunNet results, but the trial energy of the NNQS trial wave function for the related NNQS-AFQMC results. 
    MD-AFQMC results, with trial wave functions from truncated MCSCF, was taken from Ref.~\cite{AFQMC_bondBreaking}. The dashed green line represents the chemical accuracy (1\,kcal/mol).}
    \label{fig:energy_comparison}
\end{figure}

\begin{figure}[!ht]
    \centering
    \includegraphics[width=1.0\linewidth]{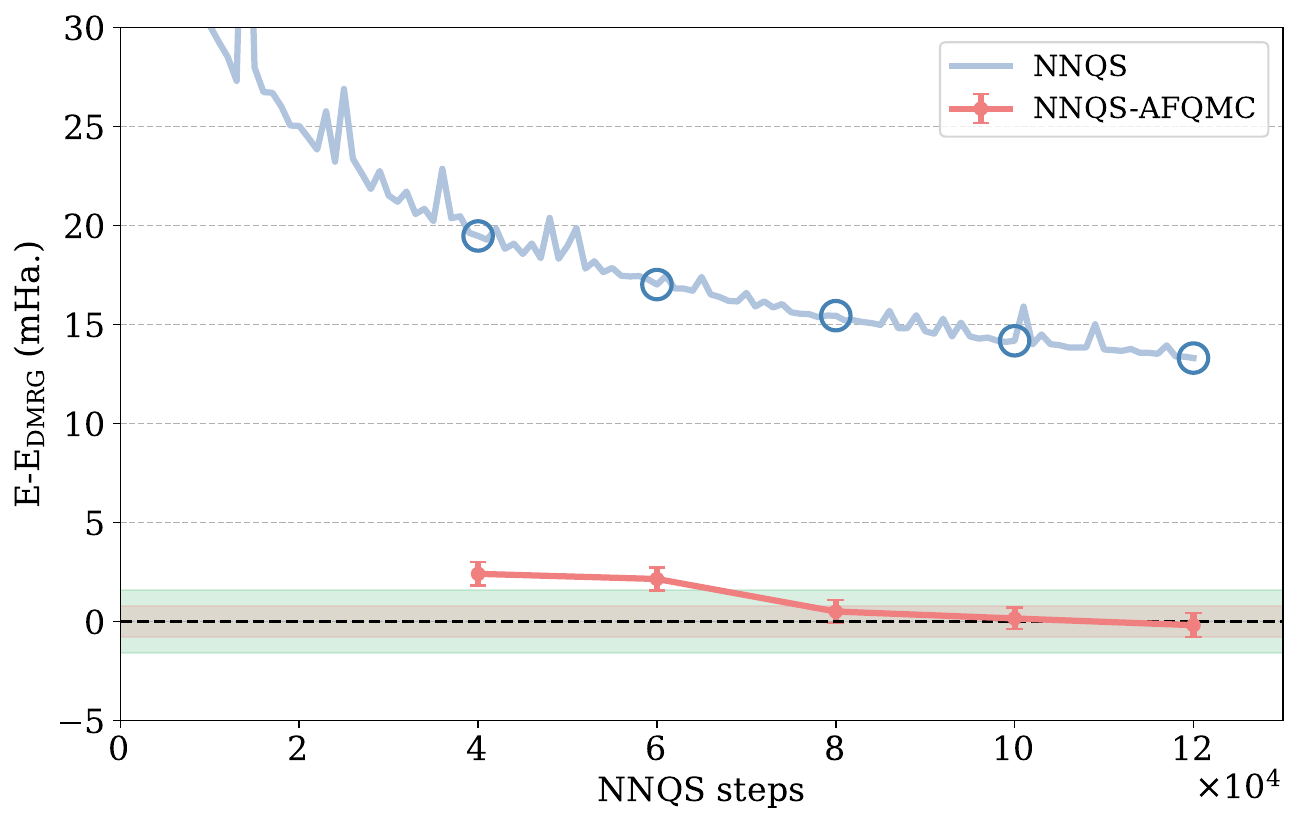}
    \caption{The convergence of NNQS-AFQMC along the training process of NNQS for $\mathrm{N}_2$ at 4.2 Bohr (cc-pVDZ basis). Blue curves indicate the NNQS energy during the training process. AFQMC is carried out with NNQS trial wave functions optimized at $4, 6, 8, 10, 12\times10^4$ steps correspondingly, denoted in the figure with blue dots. The green shaded region indicates the chemical accuracy (1 kcal/mol), and the red shaded region specifies the convergence of NNQS-AFQMC.}
    \label{fig:diff_nnqs_wf}
\end{figure}

The main results of this study are summarized in Fig. \ref{fig:energy_comparison}, highlighting the improvements of our integration. In the figure, NNQS-AFQMC is compared against AFQMC with multi-determinant trial wave functions (MD-AFQMC) in Ref.\cite{AFQMC_bondBreaking} with trial wavefunctions consisting of $O(10^3)$ configurations from MCSCF under a weight cutoff of $0.01$. \added{The NNQS energies presented in Fig. \ref{fig:energy_comparison} and the NNQS wavefunctions utilized in the NNQS-AFQMC calculations were obtained from a consistent training procedure of $12 \times 10^4$ iterations for all six points along the dissociation curve in Fig. \ref{fig:diff_nnqs_wf}.} \added{The NNQS wavefunctions utilized in the NNQS-AFQMC calculations do not represent the optimized NNQS results. Accurate NNQS results \cite{liu2025efficientoptimizationneuralnetwork} for $\mathrm{N}_2$ are reported during our preparation of this work. However, for larger or more strongly correlated systems, obtaining highly accurate NNQS wavefunctions remains challenging with current techniques, which are considered one of the main bottlenecks in the development of NNQS.
To understand the behavior of the NNQS-AFQMC algorithm with different quality of trial wave functions and pave the way for the following applications to harder systems where well-optimized NNQS trial wave functions are not available, studies with a series of non-optimized NNQS wave functions are necessary.} 

For $\mathrm{N}_2$ at the bond length of 2.118 Bohr (i.e., weakly correlated regime), both methods show excellent agreement with reference data, consistent with the weakly correlated nature of this system, where even a single Hartree-Fock solution is sufficient as the trial wavefunction. Difficulties arise as the bond is stretched and the system enters the strongly correlated regime. The Ref.\cite{AFQMC_bondBreaking} fails to maintain chemical accuracy (1 kcal/mol) for bond lengths exceeding 2.4 Bohr. NNQS-AFQMC for all bond lengths reaches near-exactness, demonstrating exceptional robustness across both equilibrium and strongly correlated regimes. The failure of MD-AFQMC in the strongly correlated regime corresponds to the exponentially increased number of configurations in MCSCF to characterize the ground state, while MD-AFQMC can only hold trial wave functions with configurations around $O(10^3)$. The improvements of NNQS-AFQMC benefit from our stochastic sampling of the trial wave function, which captures the exponential increase of configurations. Similar study of $\mathrm{N}_2$ bond breaking are also presented in Ref.~\cite{xiao2025implementingadvancedtrialwave}, where highly accurate results are achieved while using VAFQMC trial wave function and sampled in auxiliary field space. 

Our data further establishes that NNQS-AFQMC exhibits remarkable robustness to the variational accuracy of the underlying NNQS. Crucially, our study prioritizes convergence of the AFQMC energy rather than exhaustive optimization of the NNQS variational energy: even partially optimized NNQS with residual variational errors can serve as effective trial states for AFQMC. \added{In practice, the optimization of the NNQS trial wave function and related AFQMC are performed in a parallel manner. We monitor the behavior of AFQMC with different NNQS trial wave functions that are optimized with increased steps, and we stop the training of NNQS, once there are converged AFQMC results.}

As presented in Fig. \ref{fig:diff_nnqs_wf} for $\mathrm{N}_2$ at 4.2 Bohr, during the NNQS training process, five distinct trial wavefunctions are taken at training steps $4, 6, 8, 10, 12\times10^4$—capturing progressive stages of variational energy convergence. For each wavefunction, NNQS-AFQMC is performed to examine the dependence of AFQMC results on its wavefunction. At $4\times10^4$ and $6\times10^4$ training steps, NNQS energies show 15-20 mHa errors. Though NNQS-AFQMC improves upon these results (yielding $\leq3$ mHa errors), it fails to reach chemical accuracy. For $8\times10^4$ steps, where NNQS errors drop to 10-15 mHa, NNQS-AFQMC shows consistent convergence with near-exact results. Beyond $10\times10^4$ steps, AFQMC results converge while further NNQS optimization yields negligible improvements in AFQMC performance, through NNQS trial wavefunctions still show potential to reach lower energy (i.e., not converged). \added{Critically, despite ongoing optimization of the neural network quantum state (NNQS) wavefunction, further improvements in accuracy are characterized by sharply diminishing returns. Generally, the computational time per NNQS iteration remains approximately constant; hence, the total training time scales linearly with the number of steps. Achieving higher precision in the late stages of optimization demands a substantially increasing number of iterative steps.} This consistent convergence behavior is likewise observed for the remaining points shown in Fig. \ref{fig:energy_comparison}. This behavior underscores a key efficiency gain: AFQMC effectively corrects systematic errors in the trial wavefunction, rendering exhaustive NNQS training unnecessary once a minimal quality threshold is met, yielding substantial computational advantages in the training of NNQS.

One important parameter in the integration of our approach is the number of configurations, $P$, as defined in Eq.~\ref{eq:Metropolis_sampled_trial}. This parameter directly determines the efficiency of our implementation. To investigate its behavior, NNQS-AFQMC calculations are performed with different $P$ values. Representative results for $\mathrm{N}_2$ at a bond length of $4.2$ Bohr\added{, obtained using an NNQS trial wavefunction after $12\times10^4$ iterations,} are illustrated in Fig. \ref{fig:diff_nnqs_wf}. The data show that NNQS-AFQMC attains chemical accuracy when $P \geq 200$, and convergence is achieved around $P = 400$. The decrease in both the systematic energy error and statistical uncertainty as $P$ increases aligns with the findings reported in Ref.\cite{xiao2025implementingadvancedtrialwave}, underscoring a fundamental characteristic of the stochastic sampling process employed in the integration of AFQMC with many-body trial wave functions. Throughout our study of $\mathrm{N}_2$, a configuration number of $P = 400$ was sufficient to ensure convergence of all NNQS-AFQMC calculations. \added{We also observed that the choice of $P$ is almost consistent across different training steps in the study of $\mathrm{N}_2$. More specifically, in all of our tested cases, $P=400$ is enough for all NNQS trial wave functions (i.e., optimized with $4\times 10^4$ to $12\times 10^4$ steps) to obtain converged AFQMC results. } As detailed in Sec.~\ref{sec:methods_NNQS_AFQMC}, the computational cost of our NNQS-AFQMC implementation scales sub-linearly to $P$. The computational cost of NNQS-AFQMC is comparable to that of AFQMC using CAS trial wave functions, where generally $O(10^3)$ configurations can be efficiently processed \cite{ipie}. 

\begin{figure}[h]
    \centering
    \includegraphics[width=0.9\linewidth]{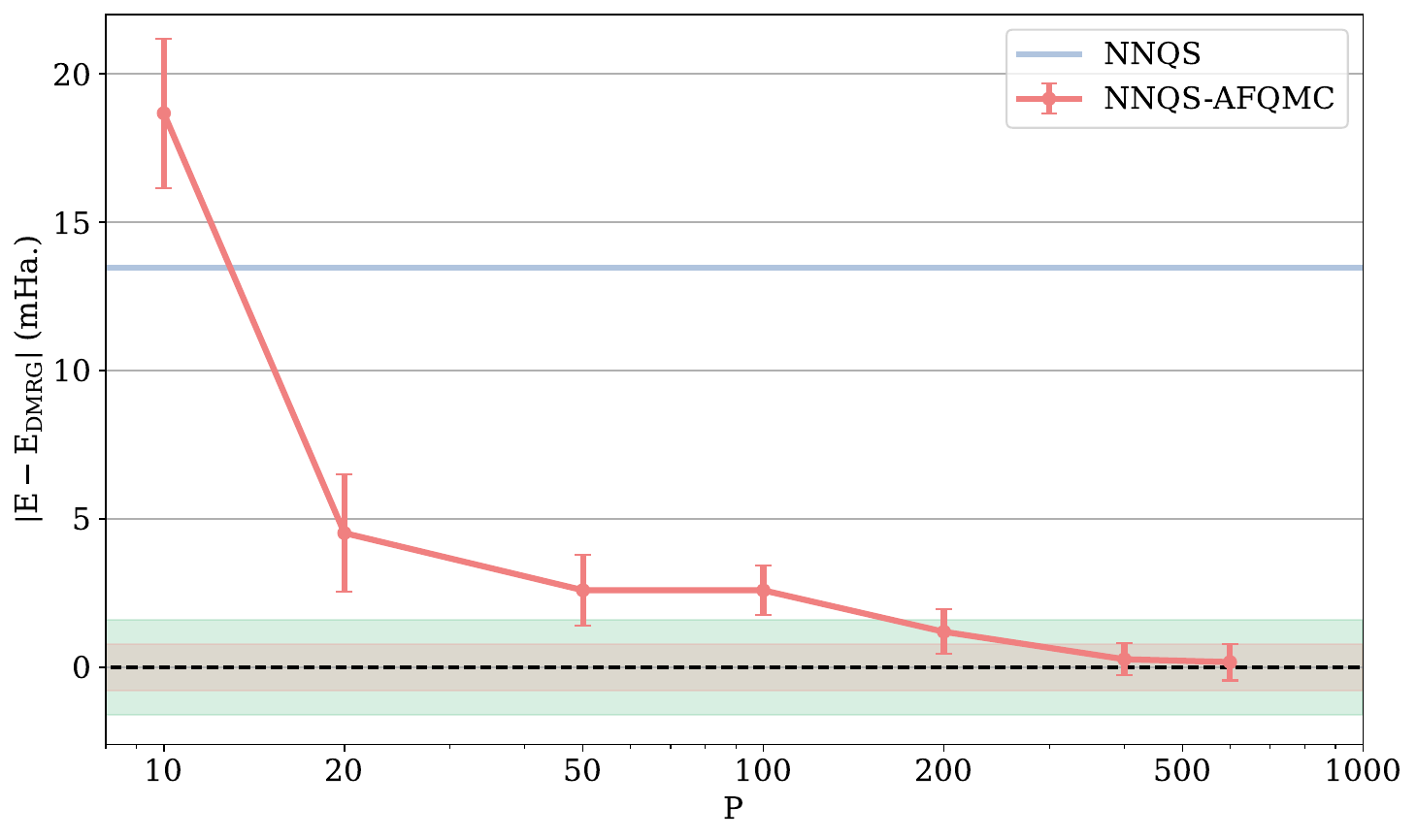}
    \caption{Energy deviation of NNQS-AFQMC vs. the number of Metropolis samples (P)  for $\mathrm{N}_2$ at 4.2 Bohr (cc-pVDZ basis). Calculations are performed with the same NNQS trial wave function and parameters, except for the number of Metropolis samples $P$. NNQS denotes the trial energy of the NNQS trial wave function. The green shaded region indicates the chemical accuracy (1 kcal/mol), and the red shaded region specifies the convergence of NNQS-AFQMC. }
    \label{fig:diff_chains}
\end{figure}

\added{
Based on the above studies, we applied NNQS-AFQMC to $\mathrm{N_2}$ at 4.2 Bohr with larger cc-pVTZ and cc-pVQZ basis sets, $\mathrm{H_2O}$ molecule with cc-pVDZ basis set, and benchmark with other accurate methods. All calculations were performed with the same setup and AFQMC parameters, as described in Sec. 4. To evaluate the sensitivity to sample size, we tested four values of the parameter \(P\): 100, 200, 400, and 800. The results, summarized in Table.~\ref{tab:ccpvtz_ccpvqz_results}, demonstrate that NNQS-AFQMC consistently achieves high accuracy across both molecular systems and basis sets using a fixed set of parameters. Notably, a sample size of \( P = 400 \) is proved sufficient to maintain chemical accuracy in all cases. Additional wall-time data are provided in APPENDIX.~\ref{sec:appendix_walltime} to further illustrate the computational performance of our approach.
}

\begin{table}
\begin{tabular}{ccccc}
\hline
\hline
Method & $\mathrm{H_2O}$(cc-pVDZ) & $\mathrm{N_2}$(cc-pVDZ) &$\mathrm{N_2}$(cc-pVTZ)&$\mathrm{N_2}$(cc-pVQZ) \\
\hline
NNQS (trial) & -76.2408& -108.9569 &  -109.0452 &  -109.0865  \\
NNQS-AFQMC P=100 & -76.2437(4) & -108.9727(8) & -109.067(1) &  -109.121(2) \\
NNQS-AFQMC P=200 & -76.2433(3) & -108.9720(7) & -109.066(1) &  -109.118(1) \\
NNQS-AFQMC P=400 & -76.2431(3) & -108.9704(5) &-109.0633(9) &  -109.119(1) \\
NNQS-AFQMC P=800 & -76.2427(2) & -108.9703(6) & -109.0634(7)
&  -109.117(1) \\
FCI       & -76.24248 & --- & --- & --- \\
DMRG\cite{alsaidi2007}      & --- & -108.97009   &    ---       &   ---    \\
CDFCI\cite{zhang2025parallel}     & --- & -108.97013   &    ---       & -109.11668 \\
UCCSD(T)\cite{alsaidi2007,mahajan2021taming}  & -76.24187 & -108.96299   & -109.05442   & -109.1117\\
UCCSDT\cite{alsaidi2007}    & --- & -108.96685   & -109.0585    &   ---   \\
QMC/MCSCF\cite{alsaidi2007} & --- & -108.9673(5) & -109.0629(7) &  ---  \\
fp-AFQMC\cite{mahajan2021taming}  & --- & -108.9702    &     ---      & -109.1192 \\
fp-AFQMC/UCCSD(T)\cite{mahajan2021taming}  & --- & ---  &     ---      &  -109.1189 \\
     \hline
     \hline
\end{tabular}
\caption{Total energies (in E$_h$) for the $\mathrm{N}_2$ and $\mathrm{H_2O}$ molecules computed using the NNQS-AFQMC method with different sample size $P$, benchmarking with other methods. The $\mathrm{N}_2$ molecule was studied at a bond length of $4.2$ Bohr with the cc-pVXZ (X = D, T, Q) basis sets, while $\mathrm{H_2O}$ was computed at O–H bond length of $0.94237 \mathrm{\AA}$ and H–O–H bond angle of 107.17 \textdegree using the cc-pVDZ basis set. FCI data is obtained with PySCF \cite{pyscf_1}. }
\label{tab:ccpvtz_ccpvqz_results}
\end{table}

\section{Summary and discussion}
\label{sec:Summary}
In this work, we introduced an integration of NNQS with AFQMC. Our approach builds upon recent progress in AFQMC\cite{xiao2025implementingadvancedtrialwave} and adapts it to the latest NNQS formalisms. Such integration leverage the variational flexibility of NNQS to construct an optimized trial wavefunction that significantly enhances the accuracy of AFQMC simulations. The integration is rigorously tested on $\mathrm{N}_2$ molecules under systematic bond stretching, spanning weakly correlated to strongly correlated regimes. Across all regimes, our method systematically yields results around exactness, showcasing its potential for tackling challenging quantum systems.

This work highlights the versatility of the framework in the original work \cite{xiao2025implementingadvancedtrialwave} and introduces a detailed implementation of stochastic sampling to more general many-body trial wavefunctions. It is noticed that, as discussed in Sec.~\ref{sec:methods_NNQS_AFQMC}, the precomputed dataset utilized for sampling does not depend on the specific form of the trial wavefunction. Once a desired pre-computed dataset is obtained, the following AFQMC algorithm can be processed with a trial wavefunction described by such pre-computed dataset. Therefore, our integration is not limited to the detailed NNQS format and can be extended to a broad class of trial wavefunctions that previously suffered from the scaling in their AFQMC implementations. Notably, this methodology seamlessly accommodates Configuration Interaction (CI) excited states, including those derived from CAS \cite{Morales_MD_AFQMC, Hao_Some_recent_developments}, SHCI \cite{SHCI_AFQMC}, and CISD \cite{Ankit_CISD}. In these cases, their CI configurations can be directly deposited as pre-computed datasets. Following stochastic sampling from such pre-computed datasets, together with carefully designed fast update algorithms, can significantly reduce computational scales compared to traditional AFQMC approaches. 

Other quantum chemical methods can also benefit from the development of AFQMC, especially VMC methods. A conceptual understanding of AFQMC here is to view it as a fair approximation of long-time imaginary time evolution $e^{-\beta \hat{H}}$ that now can be applied to many-body wavefunctions. Such AFQMC approximation of long-time imaginary time evolution is free of the sign problem, and its quality is associated with the choice of many-body trial wavefunctions. In light of these characteristics, AFQMC emerges as a robust validation framework for VMC results. By adopting VMC optimized wavefunctions as trial wavefunctions in AFQMC, researchers can conduct rigorous assessments of VMC predictions, thereby enhancing the overall reliability of the computational outcomes. Beyond validation, AFQMC can offer valuable insights into the detailed optimization processes within VMC, such as guiding the choice of variational parameters or suggesting more efficient sampling strategies. This bidirectional exchange of ideas and methodologies holds the promise of driving significant advancements in the field of quantum chemistry, enabling more accurate and efficient exploration of many-body quantum systems.

\section{Data Availability}
The code for QiankunNet and the data that support the findings of this study are available from the corresponding author upon reasonable request.

\begin{acknowledgement}
This work is supported by National Natural
Science Foundation of China (Grant No. T2222026). This work was mainly supported by the Supercomputing Center of the USTC. Part of this work was carried out with the support of Institute of Physics, Chinese Academy of Sciences.
\end{acknowledgement}

\bibliography{achemso-demo}

\appendix

\begin{appendices}
\appendixpage
\section{Details for NNQS-AFQMC}
\label{sec:APPENDIX_summary}
In this section, we elaborate on the details of implementing AFQMC with NNQS trial wavefunction through stochastic sampling, which can be divided into three main steps: initialization, propagation, and measurement.

\subsection{Initialization}
\label{sec:APPENDIX_summary_Initialization}
The precomputed dataset for NNQS is constructed below:
\begin{itemize}
\item generate CI representations of NNQS trial wavefunction. In this work, we take all configurations used in NNQS estimation.
\item sort all configurations from largest to smallest according to the absolute value of their coefficients.
\item label sorted configurations from $1$ to $N_{tot}$ correspondingly. Without loss of generality, we take $1$ to specify the configuration with the largest coefficient and  $N_{tot}$ to be the least. $N_{tot}$ indicates the total number of configurations to represent the NNQS wavefunction.
\end{itemize}  

With a precomputed dataset, we can initialize the sampled trial wavefunction
\begin{equation}
\begin{aligned}
\langle \Psi^{0}_{T,k}|
=\sum_{p=1}^P
\frac{\Psi_T(\mathbf{x}_p)\langle \mathbf{x}_p |}
{\big| \Psi_T(\mathbf{x}_p)\langle \mathbf{x}_p|\phi^0_k\rangle \big|},
\end{aligned}
\end{equation}
where $\mathbf{x}_p$ indicates the sampled labels that range from $1$ to $N_{tot}$ and $|\phi^0_k\rangle$ here is chosen to be RHF wavefunction. The sampling of $\mathbf{x}_p$ with probability 
$$
{\mathcal P}(\mathbf{x}_p;\phi) = \big| \Psi_T(\mathbf{x}_p)\langle \mathbf{x}_p|\phi\rangle \big|/{\mathcal N}(\phi)
$$
can be done by the standard Metropolis process:
\begin{itemize}
\item Initialize a set of $\{\mathbf{x}_p\}$. In our work, $\mathbf{x}_p$ is chosen to be the labels of the largest $P$ configurations.
\item Metropolis update $\mathbf{x}_p$ according to $|\phi^0_k\rangle$: 
\begin{itemize}
    \item sample $\mathbf{x}'_p$ with probability $p(\mathbf{x}_p)$. In our work, we sample $\mathbf{x}'_p$ from $[\mathbf{x}_p -P ,\mathbf{x}_p + P]$ with even probability. 
    \item accept sampled $\mathbf{x}'_p$ according to ratio 
        $$
        \alpha = \frac{ |\Psi_T(\mathbf{x}'_p)\langle \mathbf{x}'_p|\phi^{i+1}_k\rangle|}{ |\Psi_T(\mathbf{x}_p)\langle \mathbf{x}_p|\phi^{i+1}_k\rangle|}.
        $$
        If $\mathbf{x}'_p$ is accepted, set $\mathbf{x}_p=\mathbf{x}'_p$. 
    \item iterate above to reach equilibrium.
\end{itemize}   
\end{itemize}   

\subsection{Propagation}
\label{sec:APPENDIX_summary_Propagation}
Basic theories for the propagation are discussed in Sec.~\ref{sec:methods_AFQMC}. Here, we provide the implementation details:
\begin{enumerate}
    \item AFQMC advance walkers to $| \phi^{i+1}_{k}\rangle$ from $| \phi^{i}_{k}\rangle$ according to $\langle \Psi^{i}_{T,k}|$: 
    \begin{itemize}
        \item sample a field $\textbf{y}$ with probability $p(\textbf{y})$\,,
        \item evaluate dynamic shift
        $$
        \overline{\textbf{y}}^i_{k,\gamma} =-\sqrt{\tau}\frac{\langle  \Psi^i_{T,k}|\hat{L}_\gamma |\phi^i_k \rangle}{\langle \Psi^i_{T,k}| \phi^i_k\rangle}
        $$
        and
        $$
        I(\textbf{y}, \overline{\textbf{y}}^i_k, \phi^i_k) = \frac{p(\textbf{y}-\overline{\textbf{y}}^i_k)}{p(\textbf{y})}\frac{\langle \Psi^i_{T,k}| \hat{B}(\textbf{y} - \overline{\textbf{y}}^i_k)| \phi^i_k \rangle }{\langle \Psi^i_{T,k}|\phi^i_k \rangle }\,.
        $$
        \added{where $\Psi^i_{T,k}$ represents the trial wavefunction $\Psi_T$ sampled at $i$-th propagation step on $k$-th walker}
        \item advance walker
        $$
        | \phi_k^{i+1} \rangle = \hat{B}(\textbf{y} - \overline{\textbf{y}}_k^i)| \phi_k^i \rangle
        $$
        \item assign weight $W^{i+1}_k$ under the phaseless approximation:
        $$
        W_k^{i+1} = W^i_k * \bigr|I(\textbf{y}, \overline{\textbf{y}}_k^i, \phi_k^i)\bigr| * \mathrm{max}[0,\cos(\mathrm{Arg}\frac{\langle \Psi^i_{T,k}| \phi^{i+1}_k \rangle }{\langle \Psi^i_{T,k}|\phi^{i}_k \rangle})].
        $$
    \end{itemize} 
    \item Metropolis update $\mathbf{x}_p$ according to $| \phi^{i+1}_{k}\rangle$: 
    \begin{itemize}
        \item sample $\mathbf{x}'_p$ with probability $p(\mathbf{x}_p)$. In our work, we sample $\mathbf{x}'_p$ from $[\mathbf{x}_p -P ,\mathbf{x}_p + P]$ with even probability. 
        \item accept sampled $\mathbf{x}'_p$ according to ratio 
            $$
            \alpha = \frac{ |\langle\Psi_T|\mathbf{x}'_p\rangle\langle \mathbf{x}'_p|\phi^0_k\rangle|}{ |\langle\Psi_T|\mathbf{x}_p\rangle\langle \mathbf{x}_p|\phi^0_k\rangle|}.
            $$
            If $\mathbf{x}'_p$ is accepted, set $\mathbf{x}_p=\mathbf{x}'_p$ or set $\mathbf{x}'_p=\mathbf{x}_p$. 
        \item iterate above to reach equilibrium.
    \end{itemize}    
    \item Evaluate $\Delta\theta_T$ and update weight $W_k^{i+1} = W_k^{i+1}*\mathrm{max}[0,\cos(\Delta\theta_T)]$ according to Eq.~\ref{eq:AFQMC_Metro_constraint_cos}.
\end{enumerate}

\subsection{Measurement}
The measurement described in Eq.~\ref{eq:mixed_estimator} then can be evaluated with $\langle \Psi_{T}| = \langle \Psi^{n}_{T,k}|$ at step $n$ through
\begin{equation}
\begin{aligned}
\frac{\langle \Psi_T |\hat{H} |\Psi^n \rangle}{\langle \Psi_T |\Psi^n \rangle}=\frac{\sum_{k} W^{n}_k\frac{\langle \Psi^n_{T,k} |\hat{H} |\phi^{n}_k\rangle}{\langle \Psi^n_{T,k}|\phi^{n}_k\rangle}}{\sum_{k } W^{n}_k}.
\label{eq:MC_measurement}
\end{aligned}
\end{equation}
To further reduce the statistic error, we can again apply Metropolis update on auxiliary fields $\mathbf{x}_p$ in $\langle \Psi^{n}_{T,k}|$ to sample a set of new auxiliary fields $\mathbf{x}'_p$ and related $\langle \tilde{\Psi}^{n}_{T,k}|$. The measurements in Eq.~\ref{eq:MC_measurement} then can be updated through
\begin{equation}
\begin{aligned}
\langle \Psi^{n}_{T,k}| \rightarrow \langle \tilde{\Psi}^{n}_{T,k}|
\end{aligned}
\end{equation}
and
\begin{equation}
\begin{aligned}
W^{n}_k \rightarrow W^{n}_k*\frac{\sum_{p'=1}^{P'} S(\mathbf{x}'_{p'})}{\sum_{p=1}^P S(\mathbf{x}_p)}
\end{aligned}
\end{equation}
with $ S(\mathbf{x}'_{p'})$ and $S(\mathbf{x}_p)$ to specify related sign ratio. 

\section{Naive implementation of AFQMC with NNQS trial wavefunction through multi-determinants}
\label{sec:APPENDIX_MD-AFQMC}
\begin{figure}[!ht]
    \centering
    \includegraphics[width=1.0\linewidth]{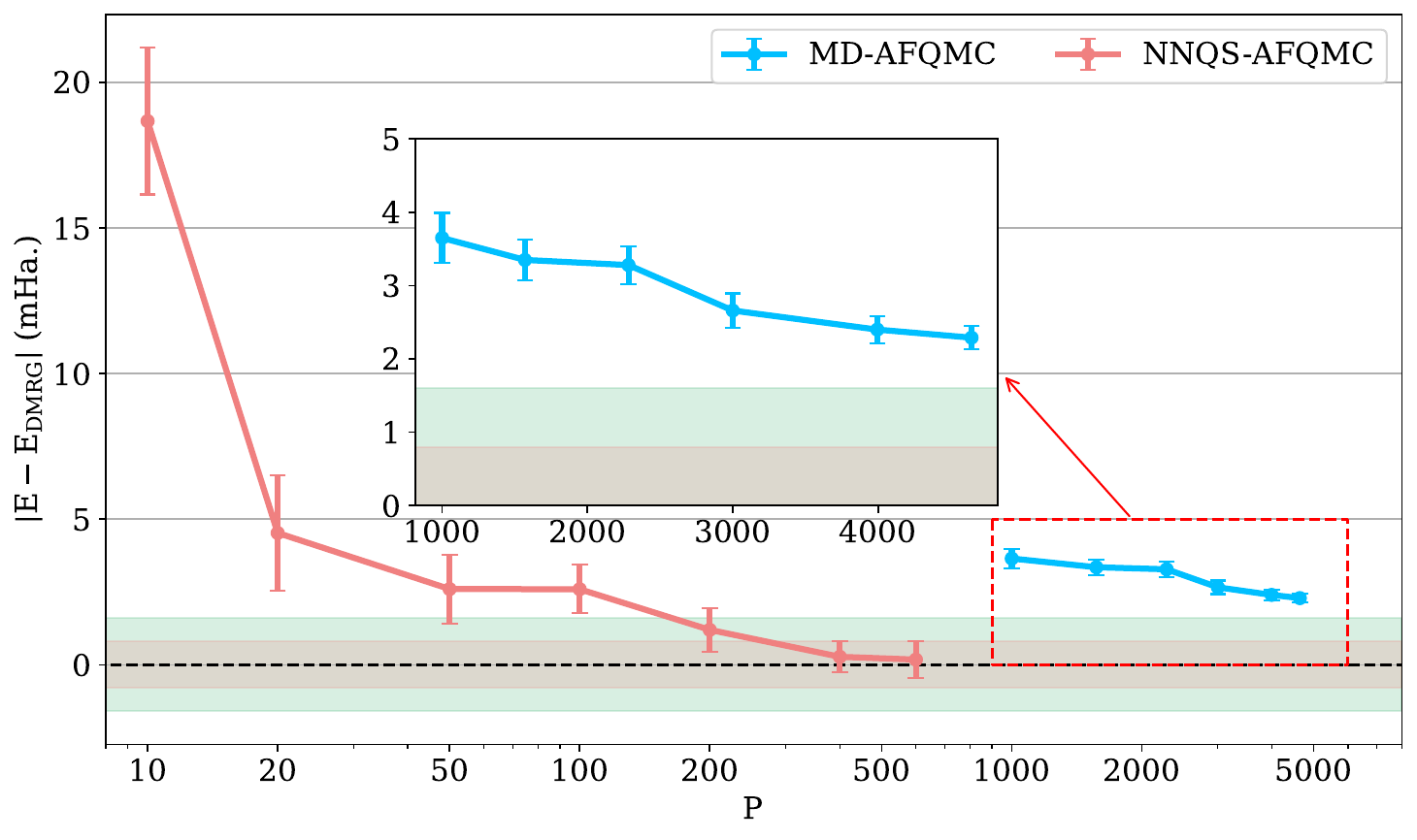}
    \caption{Comparison of NNQS-AFQMC with MD-AFQMC for $\mathrm{N}_2$ at 4.2 Bohr (cc-pVDZ basis). Both methods aim to implement the same NNQS trial wave function. NNQS-AFQMC is implemented through stochastic sampling with $P$ sampled configurations. MD-AFQMC employing multi-determinant trial wavefunctions that consist of the $P$ configurations with the largest coefficients in NNQS. The green shaded region indicates the chemical accuracy (1 kcal/mol), and the red shaded region specifies the convergence of NNQS-AFQMC.}
    \label{fig:compare_nnqsafqmc_nnqs}
\end{figure}

Fig. \ref{fig:compare_nnqsafqmc_nnqs} demonstrates the superiority of our stochastic sampling in integrating AFQMC with NNQS trial wavefunctions compared to traditional AFQMC with multi-determinant trial wavefunctions (MD-AFQMC). The figure inherits the NNQS-AFQMC results from Fig. \ref{fig:diff_chains}, supplemented with MD-AFQMC using a truncated expansion of the target NNQS. The blue curve in Fig. \ref{fig:diff_nnqs_wf} represents MD-AFQMC employing multi-determinant trial wavefunctions that consist of the $O(10^3)$ configurations with the largest coefficients from the NNQS trial wave function used in NNQS-AFQMC. Conceptually, this truncated expansion approximates the full NNQS as the number of configurations increases; however, accurately capturing the NNQS typically requires $O(10^5)$ configurations, which is computationally infeasible. The main difference between MD-AFQMC and NNQS-AFQMC here is that all configurations in the multi-determinant trial wave function are fixed during the propagation of MD-AFQMC, while NNQS-AFQMC allows these configurations to be dynamically updated during the propagation of walkers to explore all relevant configurations. Therefore, there is a persistent ``truncated error" for MD-AFQMC. Since the computational cost of these two methods shares the same scale with the number of (sampled) configurations, except NNQS-AFQMC requires an extra process to update sampled configurations. We directly compare these two methods in one figure. As presented, compared to NNQS-AFQMC which converges around $P=400$, a deviation around $2.5$ mHa is still observed in MD-AFQMC, even though $P$ is chosen to be $6\times10^3$. This comparison highlights the necessity of our stochastic sampling in integrating AFQMC with NNQS trial wavefunctions.

\section{Thermalization in the stochastic sampling of NNQS}
The stochastic implementation of the many-body trial wave function to AFQMC requires corresponding MCMC procedures to sample in equilibrium. As demonstrated in Sec.~\ref{sec:APPENDIX_summary},  thermalization to ensure equilibrium is necessary in both the initialization of AFQMC (Sec.~\ref{sec:APPENDIX_summary_Initialization}) and during the propagation of walkers (Sec.~\ref{sec:APPENDIX_summary_Propagation}). The burn-in steps (i.e., the thermalization steps to ensure equilibrium) in the initialization of AFQMC and the propagation of walkers should then be carefully investigated.

During our study of NNQS-AFQMC, we found that the burn-in steps in the initialization of AFQMC are system-dependent. For \(\mathrm{N_2}\) at stretched bond lengths (\(R = 2.7\text{--}4.2~\mathrm{Bohr}\)), \(500\) burn-in steps achieve full thermalization, yielding converged NNQS-AFQMC energies. However, for systems around equilibrium configurations (\(R_{\mathrm{eq}} = 2.118~\mathrm{Bohr}\) and \(R = 2.4~\mathrm{Bohr}\)): the minimal burn-in length increased to \(16,\!000\) steps. The substantially increased burn-in steps required for near-equilibrium geometries primarily stem from heightened probability concentration within the configuration space. At 2.118 Bohr, for instance, the Hartree-Fock (HF) state dominates the ensemble with a 93.65\% statistical weight. Such extreme probability localization severely impedes Metropolis sampling efficiency: Walkers initialized in the HF state exhibit exponentially suppressed transition probabilities to other configurations due to extremely small acceptance ratios ($\alpha \ll 1$). 

Fortunately, once equilibrium is reached in the initialization of AFQMC, the following burn-in steps in the propagation of walkers are negligible. This is because, as outlined in Sec.~\ref{sec:APPENDIX_summary_Propagation}, MCMC procedures to sampled $\langle \Psi^{i+1}_{T,k}|$ are directly updated on configurations $\mathbf{x}_p$ from the last step propagation (i.e., in sampling of $\langle \Psi^{i}_{T,k}|$), and each step of AFQMC propagation $|\phi^i_k\rangle \rightarrow |\phi^{i+1}_k\rangle$ is only separated by a sufficient small time slice $\tau$. This observation is similar to Ref.~\cite{xiao2025implementingadvancedtrialwave} where VAFQMC trial wave functions are sampled in auxiliary field space. Though an extensive burn-in step may be necessary in the initialization of AFQMC, the computational cost to reach the equilibrium of MCMC is negligible in the whole AFQMC process. 

\section{\added{Wall-times for NNQS-AFQMC calculations}}
\label{sec:appendix_walltime}

The computational efficiency of the NNQS-AFQMC method is assessed by measuring the wall-times for the systems listed in Table.\ref{tab:ccpvtz_ccpvqz_results}: $\mathrm{N_2}$ at 4.2 Bohr with larger cc-pVTZ and cc-pVQZ basis sets, $\mathrm{H_2O}$ molecule with cc-pVDZ basis set. The corresponding wall-times are summarized in Table.\ref{tab:NNQS_walltime} and Table.\ref{tab:AFQMC_walltime} respectively, for the NNQS (GPU) and AFQMC (CPU) components under various sample sizes \(P\).
\begin{table}
    \begin{tabular}{cccccc}
    \hline
    \hline
    walltime(s)&  $\mathrm{H_2O}$(cc-pVDZ) & $\mathrm{N_2}$(cc-pVDZ) &$\mathrm{N_2}$(cc-pVTZ)&$\mathrm{N_2}$(cc-pVQZ)\\
    \hline
    NNQS(GPU) & 0.57  & 0.89 & 2.62 & 12.25  \\
     \hline
     \hline
    \end{tabular}
    % \caption{Wall-times (in seconds) for one step of optimization in the NNQS components of the NNQS-AFQMC calculations. NNQS calculations are performed with one NVIDIA A800 GPU. NNQS trial wave functions used in AFQMC calculations are optimized up to $12\times 10^4$ steps. }
    \caption{Wall-times (in seconds) for one step of optimization in the NNQS components of the NNQS-AFQMC calculations. The following wall time measurements were obtained using 16 NVIDIA A800 GPUs. }
    \label{tab:NNQS_walltime}
\end{table}

\begin{table}

    \begin{tabular}{ccccccc}
    \hline
    \hline
    walltime(s)&  & $\mathrm{H_2O}$(cc-pVDZ) & $\mathrm{N_2}$(cc-pVDZ) &$\mathrm{N_2}$(cc-pVTZ)&$\mathrm{N_2}$(cc-pVQZ)\\
    \hline
    \multirow{4}{*}{AFQMC(CPU)} & P=100  & 1928  &   2865 & 8152 & 31667\\
     & P=200    & 2717  & 5690 & 15193 & 64140 \\
     & P=400    & 5367  & 11310 & 29212 & 126679 \\
     & P=800    & 14530 & 23187 & 58202 &  251995\\
     \hline
     \hline
    \end{tabular}
    \caption{Wall-times (in seconds) for the whole AFQMC components of the NNQS-AFQMC calculations. AFQMC is performed with 128 AMD EPYC 7543 CPU cores.}
    \label{tab:AFQMC_walltime}
\end{table}

\section{\added{The “sorted labeling” strategy in MCMC}}
\label{sec:appendix_sorted_labeling}

To quantitatively evaluate the performance of our proposing strategy, we compared the average acceptance ratios of two distinct proposal generation methods: (1) the “sorted labeling” strategy introduced in our paper, and (2) a random hopping strategy, where an electron is randomly annihilated from an occupied orbital and another is randomly created on an unoccupied orbital to get a new configuration. 

For both strategies, 400 independent Markov chains were initialized from the Hartree-Fock state, and each chain performed 1000 MCMC updates. The resulting average acceptance ratios, summarized in Table \ref{tab:accept_ratio}, demonstrate a markedly higher acceptance rate using the sorted labeling strategy. 

\begin{figure}
    \centering
    \includegraphics[width=0.8\linewidth]{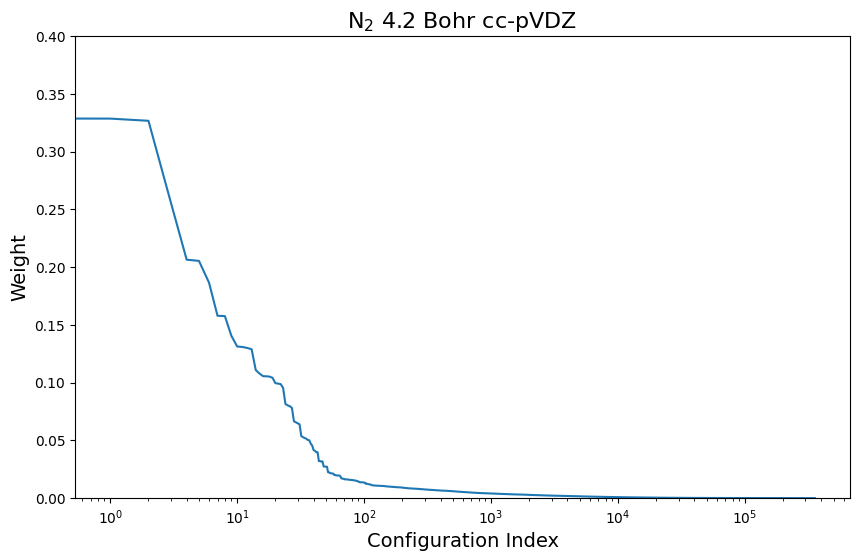}
    \caption{
    Distribution of weights for configurations in the precomputed dataset in the study of $\mathrm{N_2}$ with 4.2 Bohr bond length at cc-pVDZ basis. This precomputed dataset is obtained from the NNQS optimization at $12 \times 10^4$ steps.
    }
    \label{fig:dis_config}
\end{figure}

This improvement in the “sorted labeling” strategy benefits from the concentrated distribution of weights for each configuration in describing NNQS trial wave functions. Based on the MCMC algorithm, the acceptance of new configurations is determined by the overlap ratio between the old configuration and the newly proposed configuration. Since ``the magnitude of coefficients is associated with the importance of configurations", overlap ratios are highly relevant to the value of weights, and configurations with similar weights have a larger probability to be accepted. As demonstrated in Fig.~\ref{fig:dis_config}, the weights of NNQS are highly concentrated in configuration space. Instead of randomly proposing a new configuration, our design of the proposal process allows newly updated configurations to share similar weights with the old configuration, which significantly increases the acceptance ratios. 

\begin{table}[]
    \centering
    \begin{tabular}{cc}
    \hline
    Propose strategy     &  Acceptance ratio \\
    \hline
    Sorted labeling     &   0.143 \\
    Random hopping     &    0.00037 \\
    \hline
    \end{tabular}
    \caption{Average acceptance ratio for different propose strategies in the test cases. For both strategies, 400 independent Markov chains were initialized from the Hartree-Fock state, and each chain underwent 1000 MCMC updates with the corresponding propose strategies.}
    \label{tab:accept_ratio}
\end{table}
\end{appendices}
\end{document}